\documentstyle[preprint,eqsecnum,aps]{revtex}

\begin{document}
\draft
\preprint{}
\title{Bifurcations of a driven granular system under gravity
}
\author{Masaharu Isobe\footnote{
Present address: Department of Computational Science, Faculty of Science, Kanazawa University, 
Kakuma, Kanazawa 920-1192, Japan. Email address: isobe@cphys.s.kanazawa-u.ac.jp
}
}
\address{
Department of Physics, Kyushu University 33, Fukuoka 821-81, Japan.
}
\date{\today}
\maketitle
\begin{abstract}
Molecular dynamics study on the granular bifurcation in a simple model is presented.
The model consists of hard disks, which undergo inelastic collisions; the system is under the uniform external gravity and is driven by the heat bath.
The competition between the two effects, namely, the gravitational force and the heat bath, is carefully studied.
We found that the system shows three phases, namely, the condensed phase, locally fluidized phase, and granular turbulent phase, upon increasing the external control parameter.
We conclude that the transition from the condensed phase to the locally fluidized phase is distinguished by the existence of fluidized holes, and the transition from the locally fluidized phase to the granular turbulent phase is understood by the destabilization transition of the fluidized holes due to mutual interference.
\end{abstract}


\section{INTRODUCTION}
\label{sec:level1}
The dynamics of fluidized granular systems has attracted much attention in physics communities as a non-equilibrium statistical system\cite{hayakawa_1995,jaeger_1996,kadanoff_1999}.
Due to the fact that an elemental particle in granular systems is already macroscopic, there are two major differences in their dynamics from that of an ordinary molecular system.
First, thermal fluctuation does not play any role because relevant energy scales for the kinetic and potential energy are much larger than the thermal energy.
Second, the dynamics is dissipative because the degrees of freedom we are dealing with are coupled with microscopic processes that are not treated explicitly.
Because of these properties, granular systems require an energy source in order to be in a steady state and the external gravitational force plays an important role in their dynamics.

In most experimental situations, the systems are excited by a vibrating plate.
This is called ``Granular Vibrated Bed'', in which the effect of gravity is very large\cite{evesque_1989,clement_1992,knight_1993}.
Granular vibrated bed has been studied in many simulation works\cite{taguchi_1992,gallas_1992,aoki_1996}.
The convection\cite{hayakawa2_1995} and a linear stability analysis of surface wave\cite{bizon_1999} in the vibrated bed have been theoretically studied.
In the conventional setting of vibrated bed, the frequency of external driving plays an important role and this gives variety of patterns phases.
Experiments on pattern formation in granular vibrated bed, have demonstrated the existence of subharmonic stripe, square, and hexagonal patterns\cite{melo_1995}, as well as localized structures called oscillons\cite{umbanhowar_1996}, when the frequency and amplitude of the vibration are varied.
Comparison between kinetic theory and simulation for pattern formation has been done\cite{shattuck_1999}.
The patterns are similar in spatial structure to those observed in fluid dynamical systems, most notably parallel convection rolls in a thin layer of fluid heated from below (Rayleigh-B\'enard convection) and standing surface waves in a vertically vibrated liquid layer (Faraday instability).

However, the vibration complicates the situation.
Since energy input from the vibration depends on both the amplitude $a$ and the angular frequency $\omega$, the accurate quantities of input energy cannot be estimated from the external control parameter $\Gamma =a\omega^2/g$, which is usually used as the control parameter in the granular vibrated bed.
Therefore, it is difficult to construct the theory for the dissipative particles system in the non-equilibrium steady state.
From the numerical point of view, the Distinct Element Method (DEM)\cite{cundall_1979}, 
which is the soft-core particles model, has been often used for the studies of granular vibrated bed.
However, if we want to obtain the data for the statistical properties during the simulation, the sampling data must be taken at the same phase of external vibration cycle.
Moreover, in the dynamics of dissipative particles, the effect of different time step is not negligible.
This is a crucial effect to the particle behavior especially when the system is in the high density.
Therefore, DEM seems to have the practical disadvantages to study the correct statistical properties and dynamical behavior in the dissipative system, and is not appropriate as the simplest model for the study of the non-equilibrium steady state. 
The granular vibrated bed has these disadvantages to study the statistical characters of the granular system from the aspects of both theory and simulation.

Although there already have been many studies on the granular system for fluidization, most of these studies were a collection of the various phenomena from the experimental and numerical results, and these were discussed individually for each study.
Therefore, it becomes difficult to capture what the intrinsic characters of the granular system are.
Since the study of granular system requires new theoretical ideas beyond ordinary standard statistical mechanics or hydrodynamics, a simple ideal model system to compare the theoretical studies for the non-equilibrium steady state is required for the progress.
Recently, a series of works to find some universal frameworks have been appeared (See, Sec.~\ref{sec:2-A}).
In their study, the granular particles are regarded as a collection of inelastic hard spheres that interact with each other via hard sphere potential.
The assumption of an ideally hard sphere potential is also used in kinetic theory and the Boltzmann or Enskog approaches, which facilitates comparisons between theory and simulation in the elastic limit. 
From the numerical point of view, even if the typical distance between the particles is short, 
the molecular dynamics event-driven simulation, which is used in the hard-spheres system, is quite efficient and has smaller numerical error compared with time-step-driven method (e.g., DEM).

Since the gravitational force should have a large effect on clustering, it would be of interest to observe how the system behavior changes as the gravity sets in.
The main purpose of our study is to examine the statistical and dynamical characters of the inelastic hard disk system in the non-equilibrium steady state under gravity, and to provide extensive numerical results for theoretical studies to construct a base of the universal framework.
In this paper, we particularly focus on the competition between the effect of external driving and that of the gravitational field using the event-driven molecular dynamics simulation, systematically.

In Sec.~\ref{sec:2-A}, we will present a short review of a series of previous works using the inelastic hard sphere model with the heat bath in the non-equilibrium steady states. 
Our model is described in Sec.~\ref{sec:2-B}, and we clarify what the advantages and disadvantages of our model are in Sec.~\ref{sec:2-C} and Sec.~\ref{sec:2-D}, respectively.
In Sec.~\ref{sec:3-0}, results of numerical simulations to our model are presented.
Discussion on the characteristics of the bifurcations are described in Sec.~\ref{sec:4-0}.
Finally, we summarize our work in Sec.~\ref{sec:5-0}.

\section{MODEL}

\subsection{Non-equilibrium steady states}
\label{sec:2-A}
To achieve the Non-Equilibrium Steady State (NESS) in the granular system, energy must be injected into the system from outside.
Previous studies for NESS in the inelastic hard sphere (IHS) system may be categorized by the following six groups.
Firstly, we categorize the models by their dimensionality (1D or 2D).
Secondly, we consider the way in which the system is linked to the heat bath (i.e., heated from the boundaries/bulk).
Finally, we classify the existence or non-existence of gravity field.
We will briefly review these models for each category.

\subsubsection{1d model heated from the boundaries without gravity}
\label{sec:2-A-1}
Du et al.~\cite{du_1995} firstly studied the dynamics of quasi-elastic particles constrained to move on a line with energy input from boundaries. 
They showed that the system violates equi-partition of energy and the simple hydrodynamic approach fails to give a correct description of the system.
Similar models in NESS were also studied from the theoretical point of view\cite{grossman2_1996,vamos_1997,geisshirt_1998,goldshtein_1999}.
 
\subsubsection{1d model heated from the bulk without gravity}
\label{sec:2-A-2}
Some papers investigated the properties of the IHS fluid that is heated uniformly by random force so that it reaches a spatially homogeneous steady state.
This way of forcing was proposed by Williams and MacKintosh\cite{williams_1996} for inelastic particles moving on a line. 
Giese and Zippelius\cite{giese_1996} regarded the internal degree of vibrating rods as a thermal bath. 
They calculated the restitution coefficient $r$ as a function of the relative length of the colliding rods, the center of mass velocity, and the degree of excitation of the internal vibrations.
Puglisi et al.\cite{puglisi_1998} examined a model consisting of $N$ particles on a ring of length $L$ with the Langevin force.
They found two regimes: when the typical relaxation time $\tau^{P}$ of the driving Brownian process is small compared with the mean collision time 
$\tau^{P}_c$ the spatial density is nearly homogeneous and the probability distribution function (PDF) of velocity is Gaussian, but in the opposite limit $\tau^{P} \gg \tau_c$ one has strong spatial clustering, with a fractal distribution of particles, and the velocity PDF strongly deviates from the Gaussian.
 
\subsubsection{1d model heated from the boundaries under gravity}
\label{sec:2-A-3}
Kurtze and Hong\cite{kurtze_1998} studied the effect of dissipation on the density profile of 1D granular gas under gravity.
Perturbation analysis of the Boltzmann equation reveals that the correction of the density profile along the height $y$ due to dissipation resulting from inelastic collisions is positive for $0 \leq y < y_c$ and negative for $y > y_c$.
They also obtained the expression for $y_c$.
Using Boltzmann equation and molecular-dynamics simulations, Ram\'{\i}rez and Cordero\cite{ramirez_1999} studied the 1D column of $N$ inelastic point particles, in the quasielastic limit, under gravity.
They showed that the physical properties (e.g., density and temperature profile) in the hydrodynamic limit ($N\rightarrow \infty$) depend solely on the product $(1-r)N$, where $r$ is the restitution coefficient, which was confirmed by both theory and simulation.

\subsubsection{2d model heated from the boundaries without gravity}
\label{sec:2-A-4}
Esipov and P\"oschel\cite{esipov_1997} studied the kinetic energy distribution function satisfying the Boltzmann equation and construct the phase diagram of granular systems on the 2D particle system in a circle wall at constant temperature.
Their study has suggested that the granular hydrodynamics is valid for simple flows or for any flows in dilute and sufficiently elastic systems.
Grossman et al.\cite{grossman_1997} studied the steady state theoretically.
Their predictions compared well with numerical simulations in the nearly elastic limit.
Although the system is no longer close to equilibrium even in the low-density limit, the scaling behavior of the velocity distributions showed that a hydrodynamic treatment could still be useful.
The steady state of a low-density IHS gas confined between two parallel walls with same temperature is studied by Brey and Cubero\cite{brey_1998}.
Soto et al.\cite{soto_1999} also has studied the heat conduction between two parallel plates by molecular dynamics simulation of the IHS model.
Their results show that Fourier's law is not valid and a new term proportional to the density gradient must be added.

\subsubsection{2d model heated from the bulk without gravity}
\label{sec:2-A-5}
Peng and Ohta\cite{peng_1998} studied the 2D granular system where an external driving force is applied to each particle in the system in such a way that the system is driven into a steady state by balancing the energy input and the dissipation due to inelastic collisions between particles.
Experimentally, steady state of clustering, long-range order, and inelastic collapse are observed in vertically shaken granular monolayer, which is correspond to the system heated from the bulk in the limit of the frequency $\omega \rightarrow \infty$~\cite{olafsen_1998}.
In 2D the model may be considered to describe the dynamics of disks moving on an air table.
A similar IHS model with random external accelerations has been studied by Bizon et al.\cite{bizon2_1999} to test continuum theories for vertically vibrated layers of granular material.
To investigate the long-scale correlation and structure factor, van Noije et al.\cite{noije_1999} present a theory of the randomly driven IHS fluid in 2D simulation, and characterized its non-equilibrium steady state. 
As another way to realize the steady state, Komatsu\cite{komatsu_2000} introduced a new observation frame with a rescaled time.
This operation is the same as the velocity scaling method, which is often used as the temperature control method of the molecular dynamics simulation in equilibrium. 
He found that a phase transition with symmetry breaking, which is the circulation of the particles in the box with elastic walls, occurs when the magnitude of dissipation is greater than a critical value.

\subsubsection{2d model heated from the boundaries under gravity}
\label{sec:2-A-6}
In 2D system under the gravity, Luding et al.\cite{luding_1994} compared the results of DEM and IHS model.
The system of their study is driven by vibrations like a granular vibrated bed.
The results of DEM simulation were good agreement with that of event-driven simulation in the fluidized regime. 
They also found that the results of DEM simulation depend strongly on the time $t_c$ during which the particles are in contact.
Although there exist only a few recent laboratory experiments for 2D system under the gravity, an experiment on the stainless steel particle system in two dimensions was studied by Kudrolli et al.\cite{kudrolli_1997}.
The system was driven by the periodically moving wall and slightly inclined to apply the weak gravity.
This experiment showed that the clustering occurs around the opposite side of driving wall when there is no gravity, but the cluster migrates downward when the system is inclined even by a very small angle.

\subsection{IHSHG model}
\label{sec:2-B}
To study the statistical properties of the granular system under an external field with energy input in NESS, we propose a simple model system on the 2D inelastic hard disk system heated from the boundaries under the gravity\cite{isobe_jpsj}.

The system considered here consists of inelastic hard disks of mass $m$ and diameter $d$ in 2D space under uniform gravity.
We employ the periodic boundary condition in the horizontal direction.
The particles collide with each other with a restitution coefficient $r$.
All disks are identical, namely the system is monodispersed.
For simplicity, we neglect the rotational degree of freedom.
Because of the hard-core interaction, collision is instantaneous and only binary collisions occur.
When two disks, $i$ and $j$, with respective velocities ${\bf v}_i$ and ${\bf v}_j$ collide, the velocities after the collision, ${\bf v}_i'$ and ${\bf v}_j'$, are given by

\begin{eqnarray}
{\bf v}_i' & = &
{\bf v}_i-\frac{1}{2}(1+r)[{\bf n}\cdot ({\bf v}_i-{\bf v}_j)]{\bf n}
\\
{\bf v}_j' & = &
{\bf v}_j+\frac{1}{2}(1+r)[{\bf n}\cdot ({\bf v}_i-{\bf v}_j)]{\bf n},
\end{eqnarray}
       
\noindent
where ${\bf n}$ is the unit vector parallel to the relative position of the two colliding particles in contact.
Between the colliding events, the particles undergo free fall motion with the gravitational acceleration ${\bf g}\equiv (0,-g)$ following parabolic trajectories.
The system is driven by a heat bath with temperature $T_w$ at the bottom of the system; a disk hitting the bottom bounces back with velocity ${\bf v}=(v_x,v_y)$ chosen randomly from the probability distributions $\phi_x(v_x)$ and $\phi_y(v_y)$\cite{tehver_1998};

\begin{eqnarray}
\phi_x(v_x) & = &
\sqrt{\frac{m}{2\pi k_B T_w}} \exp{(-\frac{mv_x^2}{2k_B T_w})}~ (-\infty < v_x < \infty)
\label{eqn:heat1}
\\
\phi_y(v_y) & = &
\frac{m}{k_B T_w} v_y \exp{(-\frac{mv_y^2}{2k_B T_w})}~\quad (0<v_y<\infty) ,
\label{eqn:heat2}
\end{eqnarray}

\noindent
where $k_B$ is the Boltzmann constant.
Note that the wall temperature $T_w$ here is just a parameter to characterize the external driving and is not related to the thermal fluctuation.
We call this system ``Inelastic Hard disk System with a Heat bath under weak Gravity'' (IHSHG)\cite{isobe_jpsj} to emphasize the importance of the competition between excitation by the heat bath and the gravity.

\subsection{Advantages of IHSHG model}
\label{sec:2-C}
The IHSHG model is so simple that the system is completely characterized with only four dimensionless parameters: the restitution coefficient $r$, the driving intensity $\Lambda \equiv k_BT_w/mgd$, the system width in the unit of disk diameter $N_w\equiv L_x/d$, and the particle number of the initial layer thickness $N_h\equiv N/N_w$, where $L_x$ is the system width and $N$ is the total number of disks.

Now we compare our IHSHG model with other typical dissipative system in NESS.
Rayleigh-B\'enard(RB) convection is the typical continuous fluid system, where the energy source is the heat bath at the bottom, the dynamics is described by Navier-Stokes equation, and the system is driven by the difference of temperature between the top and the bottom.
The dissipation comes from the viscosity of fluid.
In this system, the control parameters are well known as Rayleigh Number and Prandtl Number.
There are some previous MD studies of RB convection using elastic hard disk\cite{mareschal_1988,rapaport_1988}.

The ordinary granular vibrated beds by using DEM\cite{taguchi_1992,gallas_1992,aoki_1996} are the discrete soft particles systems; the energy source is the vibration of the bottom, and the dynamics obeys Newton's equation of motion under the gravitational field.
The dissipation appears between inelastic particles, and the control parameter is $\Gamma=a\omega^2/g$, where $a$ and $\omega$ are the amplitude and the angular frequency respectively.
Recently, Aoki and Akiyama\cite{aoki_1998} proposed that the state of granular convection involving multipairs of convection rolls is governed by a dimensionless parameter ${\cal A}=\frac{\Gamma}{D/\alpha}=\frac{a\omega^2\alpha}{gD}$, where $D$ and $\alpha$ are the self-diffusion constant and the energy diffusivity $\alpha$, respectively.

Our IHSHG model is the discrete hard disks system, the energy source is the heat bath at the bottom and the dynamics obeys Newton's equation of motion under the gravitational field.
The dissipation comes from inelastic collision of particles, and the control parameter is $\Lambda = k_B T_w /mgd $.
The present system is analogous to the ordinary granular vibrated bed but simpler because it does not have an external time scale.
Since IHSHG model has no external periodic cycles and there are only four dimensionless parameters in the system, IHSHG model is easy to deal with numerical analysis and theoretical approach.

As mentioned briefly in Sec.~\ref{sec:2-A-6}, there is a recent experiment\cite{kudrolli_1997} on the stainless steel particle system in 2D similar to IHSHG model.
Based on these experimental results, we firstly reproduced the clustering phenomenon and the behaviors of NESS to simulate IHSHG model.
Consequently, we found that IHSHG model could reproduce Kudorolli's experiment qualitatively.
Note that IHSHG model might be categorized in 2D model heated from the boundaries under gravity described in Sec.~\ref{sec:2-A-6}.

\subsection{Some problems of IHSHG model}
\label{sec:2-D}
Although it is simple and easy to deal with, there are some problems in IHSHG model.
Since IHSHG model is based on the inelastic hard-sphere model, all problems come from a reason that the potential between two colliding particles is unusually stiff.
The instantaneous collisions imply that an interaction is zero when the particles are not in contact and suddenly becomes infinite when the distance between the colliding particles is zero.
Therefore, the momentum exchange occurs instantaneously.
However, in a real system, the situation is different, each contact takes a finite length of time and the force is large but finite.
The infinitely stiff hard-sphere interaction is an idealization or simplification of smooth repulsive pair-potential.

When the restitution coefficient is low (i.e., a dissipation rate is high) or kinetic energy is relatively smaller than potential energy under external gravity force, the number of collisions during a finite time often diverges.
This is called ``Inelastic Collapse''. 
It was firstly discovered while studying the 1D model system of dissipative particles with a elastic wall\cite{bernu_1990,mcnamara_1992}, thereafter in 2D\cite{mcnamara_1994}.
Although the inelastic collapse is thought as artificial pathology, several proposals exist to avoid it\cite{giese_1996,luding2_1994,deltour_1997,mcnamara_1996,brilliantov_1996,goldman_1998,luding_1998}.
The velocity dependence of the restitution coefficient was experimentally measured for binary collisions, and the quantitative study of the internal modes of a granular particle is under way\cite{gerl_1999}. 
However, the inelastic collapse has never appeared when the dissipation is low and the system is highly excited.
Therefore, IHSHG model for the parameter region studied here does not need to take care of it.

Since the dynamics proceeds by only the event of binary collision, multiparticle interactions are impossible to realize.
In a real system, each contact takes a finite time so that multiparticle contacts are possible.
Nevertheless, in the fluidization study, multiparticle collision can be thought as a rare event.
Hence, we can ignore this effect.

Another problem with the inelastic hard-sphere model is that the static limit does not exist. 
For example, in the framework of the inelastic hard-sphere model, a particle cannot rest motionless on the ground.
In the realistic granular systems, the potential and elastic energies will approach constant values while the kinetic energy tends to zero.
If the particles would be dissipative hard particles, the inelastic collapse will occur before the kinetic energy vanishes. 
The contact forces and stresses are not properly defined so that the system will never reach a static configuration.
However, since IHSHG model has a heat bath at the bottom of the system, the kinetic energy always stays finite.
In the simulation of NESS, the dynamics has little effect from this problem.

\section{NUMERICAL SIMULATIONS}
\label{sec:3-0}
In this section, we present the results of numerical simulation on IHSHG model using the simple and efficient event-driven algorithm\cite{isobe_ijmpc}, which allowed us to study the statistical and dynamical properties for a wide range of parameters. 

\subsection{Numerical setting}
\label{sec:3-A}
Most of the simulations were performed on the system with $r=0.9$, $N_w=200$, and $N_h=20$ (i.e., $N=4,000$).
The driving intensity $\Lambda \equiv k_BT/mgd$ was varied from 25.0 to 909.1 to study the change of the system behavior.
In the actual simulation, we reduced any variables to dimensionless ones by the following three basic units ($k_BT, m, d$), which are all fixed at unity.
Note that unit time $\tau (\tau=\sqrt{\frac{md^2}{k_BT}})$ is also unity in these simulations.
To ensure that the system had reached a steady state, we relaxed the system until the total energy did not drift. 
Therefore, the system reaches the non-equilibrium steady state.
Thereafter, we started the simulations with the length of 100,000 collisions per particle; we could obtain the data with small statistical error.
In addition, we performed the simulations with $N_w=400$ (i.e., $N=8,000$) and $N_w=1,000$ (i.e., $N=20,000$) to study the size-dependence of the system behavior. 
These run lengths were 30,000 and 10,000 collisions per particle, respectively.
The discussion for the size-dependence of the system is presented in Sec.~\ref{sec:4-0}.

\subsection{System behavior upon changing external driving}
\label{sec:3-B}

\subsubsection{Packing fraction and excitation ratio}
\label{sec:3-B-1}
In order to characterize the steady states of the system, firstly, we measure the maximum packing fraction $A_{\rm max}$ at the height $y=H_{Amax}$ as a function of the driving intensity $\Lambda$. 
In the system with $(r,N_w,N_h)=(0.9,200,20)$, there are two cusps around $\Lambda \sim 200$ and $380$($\circ$ in Fig.~\ref{fig:1}), suggesting phase transitions with changing $\Lambda$.
These transitions should be related to the excitation structure of the state, therefore, we define the excitation ratio $\mu$ by $\mu\equiv{\cal K}/{\cal U}$, namely, the ratio of the total kinetic energy ${\cal K}=(m/2)\sum_i v_i^2$ to the potential energy ${\cal U}=mg\sum_i y_i$.
The cusps appear at the same points for $\mu$, which confirms that some transitions occur at these points ($\times$ in Fig.~\ref{fig:1}).

\subsubsection{Snapshots}
\label{sec:3-B-2}
In order to conceptualize the underlying physical mechanism of the system behavior in each phase separated by the transitions, three snapshots for typical values of $\Lambda$ for each phase are shown in Fig.~\ref{fig:2}, in which interval time between them are all $100 \tau$.

For $\Lambda=181.8$ (a1---a3 in Fig.~\ref{fig:2}), most of the particles are aggregated around the bottom of the system with an almost close packed density, and the state of a dense layer is relatively stable, which means the potential energy is dominant and the system is in weakly excited states.
We call the phase in $\Lambda\leq 188.7$, the condensed phase (CP).
It can be seen from the snapshots, however, that collective motion appears on the surface of the layer.

In the second phase, $\Lambda=250.0$ (b1---b3 in Fig.~\ref{fig:2}), the dense layer packing is locally broken by excitation of the heat bath.
The high-speed particles are blown upward from the holes in the layer.
In Fig.~\ref{fig:2}, we see only one hole in the system, but for a larger system, there are certain cases where we can observe more than one hole.
The holes migrate and occasionally become more active; sometimes they almost close temporarily, but the structure is fairly stable.
We call this phase the locally fluidized phase (LFP).

For the case of $\Lambda=666.7$ in the third phase (c1---c3 in Fig.~\ref{fig:2}), the layer is completely destroyed.
The average density is quite low, but it is very different from the ordinary molecular gas phase.
The density fluctuation is large and this fluctuation causes turbulent motion driven by the gravity.
At some time, the whole system gets excited with some relatively smaller density fluctuations, but the very next moment, a large proportion of the particles travels downward and forms a layer like structure.
This structure, however, is destroyed immediately.
We call this phase the granular turbulent phase (GTP).

\subsubsection{Collision rate}
\label{sec:3-B-3}
Since the simulation was performed in the non-equilibrium steady state, the collision rate takes constant value during the whole run. 
Therefore, when the total number of collision $C_{total}(=100,000\times N)$ was fixed, the total time $T_{total}$ for each run were different for each $\Lambda$.
We found that the system in the non-equilibrium steady state can be also characterized by the collision rate because it seems to correspond to the pressure of molecular system in the equilibrium. 
Though in the elastic hard disk system the collision rate is only the function of packing fraction of the system, in the inelastic hard disk system it depend on the competition between the inelasticity $r$ and the external driving $\Lambda$.

In Fig.~\ref{fig:3}, the collision rate ($C_{total}/T_{total}$) vs. the external driving $\Lambda$ in the log-linear scale is shown.
We found that the collision rate decrease exponentially in LFP and GTP ($\Lambda > 188.7$) with different exponents.
However, the collision rate deviates from exponential fitting in CP.
It diverges when the system is in the weakly excited state ($\Lambda \rightarrow 0$).

\subsubsection{Probability distribution function of velocity}
\label{sec:3-B-4}
These inhomogeneous behaviors should result in a non-Maxwell-Boltzmann distribution of velocity.
Figure~\ref{fig:4} shows the horizontal velocity Probability Distribution Functions (PDF) at the horizontal layer around the height of the maximum packing fraction ($y=H_{Amax}$) for each phase in the log-linear scale.
The data are plotted in the unit of $\sqrt{k_BT/m}$.
The distributions for $\Lambda=250.0$ and $666.7$ deviate from the Gaussian and are more or less in exponential form in the tail region.
It can be seen that the distribution for $\Lambda=250.0$ in LFP consists of two parts; the central part that originates from the dense layer packing, and the wide tail that comes from the fluidized holes.
The central parts for all cases are very close to Gaussian (inset of Fig.~\ref{fig:4}).

\subsubsection{Flatness parameter}
\label{sec:3-B-5}
In order to quantify the deviation from Gaussian distribution, we next calculated the flatness parameter $f$ as a function of the height $y$ defined by
\begin{equation}
f(y)\equiv \langle v_x^4 \rangle /\langle v_x^2\rangle^2,
\end{equation}

\noindent
which is often used in the study of the intermittency of the fluid turbulence.
The dependence of the external driving of the maximum flatness $f(H_{fmax})$ at the horizontal layers with the thickness of disk diameter $d$ is shown in Fig.~\ref{fig:5}.
Over the whole region of external driving $\Lambda$, the value of $f(H_{fmax})$ is different from 3, which correspond to the value for the Gaussian distribution, but it is remarkable that $f$ becomes very large, as large as 20, in LFP.
In Fig.~\ref{fig:5}, the statistical error bars are also plotted for each data.
Note that the statistical errors of the time averages for molecular dynamics simulation were estimated by the block average method described in ref.\cite{allen_1987}. 

\subsubsection{Fitting functions of velocity PDF}
\label{sec:3-B-6}
The unexpected large value of $f$, however, can be understood as follows~\cite{isobe_jpsj}.
Assume the velocity PDF $\phi(v)$ has two components, the narrow Gaussian distribution $\phi_{G}(v)$ with the weight~$1-p$ and the broader stretched exponential distribution $\phi_{S}(v)$ with the weight~$p$;

\begin{equation}
\phi(v) = p \phi_{S}(v) + (1-p) \phi_{G}(v),
\quad (0\le p\le 1),
\label{eqn:phi}
\end{equation}

\noindent
where $\phi_{S}(x)$ and $\phi_{G}(x)$ are defined by
$\phi_S(x)\equiv (\beta/(2x_0\Gamma(1/\beta)))\exp{(-{|x/x_0|}^\beta)}$
and $\phi_G(x)\equiv (1/(\sqrt{2\pi}\sigma_G))\exp{(-x^2/(2\sigma_G^2))}$, respectively.
Then, the flatness of this distribution is given by
\begin{equation}
f = { [\Gamma(5/\beta)/\Gamma(1/\beta)]\tilde x_0^4 p +3(1-p)
\over \bigl(
[\Gamma(3/\beta)/\Gamma(1/\beta)]\tilde x_0^2 p+ (1-p)
\bigr)^2};
\quad
\tilde x_0 \equiv {x_0\over\sigma_G}
\label{eqn:flatness}
\end{equation}

\noindent
because the second and the fourth moment of $\phi_S(x)$ are given by $\Gamma(3/\beta)/\Gamma(1/\beta)\cdot x_0^2$ and $\Gamma(5/\beta)/\Gamma(1/\beta)\cdot x_0^4$, respectively.
If we fit the parameters in Eq.~(\ref{eqn:phi}) to the data for $\Lambda=250.0$ for example, we obtain $\sigma_G=0.048, \tilde x_0=1.7$, $\beta=0.85$, and $p=0.25$, which yields $f\simeq 19.0$ (Fig.~\ref{fig:6}).
The enhancement of the flatness by the superposition of the broader distribution can even be drastic for a small value of $p$ as can be seen if Eq.~(\ref{eqn:flatness}) is plotted as a function of $p$.
This observation indicates that $f$ can be used as a sensitive index to detect the appearance of a small weight of the broad component in the distribution.
The sharp rise of $f$ in Fig.~\ref{fig:5} around $\Lambda \sim 200$ is clear evidence of the appearance of the fluidized holes.

The other PDFs for each $\Lambda$ in LFP and GTP are also fitted by changing $p$
(e.g., $\Lambda=476.2$ in Fig.~\ref{fig:6}). 
From these $p$, we obtain the plot of $p$ for each $\Lambda$ in Fig.~\ref{fig:7}.
Near CP-LFP critical point $\Lambda_{CL}$, the weight of $p$ takes a certain finite value, not $0$ (Fig.~\ref{fig:7}).
In LFP, the weight $p$ increases linearly when $\Lambda$ increases, which means the fluidized holes region increases linearly.
However, there is a plateau $p \sim 0.5$ when the transition from LFP to GTP occurs.

\subsection{Height dependencies of flatness and packing fraction}
\label{sec:3-C}
The height dependencies of the packing fraction $A(y)$ and the flatness $f(y)$ are calculated in the horizontal layers with the thickness of the disk diameter $d$.
In Fig.~\ref{fig:8}, we plot the $\Lambda$ dependencies of $H_{fmax}$ and $H_{Amax}$, which are the heights where the flatness and the packing fraction take their maximum values.
In LFP and GTP, the both of the parameters reach their maximum values at the almost same height ($H_{fmax} \simeq H_{Amax}$), but in CP the flatness shows its maximum value at the lower position than the packing fraction does ($H_{fmax} < H_{Amax}$).
In CP, the value of the packing fraction where the flatness takes its maximum value is very close to the value of the Alder transition point $A_c$($\sim 0.7$) in the elastic hard disk system (Fig.~\ref{fig:9}).
The fact is somewhat intriguing, but we could not find obvious reason for it.

In the simulation of the granular vibrated bed, Taguchi and Takayasu\cite{taguchi_1995} showed that velocity PDF becomes power distribution in the fluidized region, but Gaussian in the solid (close packed) region.
On the other hand, Murayama and Sano\cite{murayama_1998} found that the flatness varied from Gaussian ($f=3$) to exponential ($f=6$) when the packing fraction increases.
These results seem to disagree with each other.
However, in CP of IHSHG model, we obtained two different characteristics about velocity PDF; (i) The region $0<y<H_{fmax}$, where the packing fraction increases from $0$ to $A_c$ and the flatness also increases, (ii) the region $H_{fmax}<y<H_{Amax}$, where the packing fraction $(>A_c)$ reaches nearly the closed packing and the flatness decreases.

\subsection{Spatial velocity correlation function}
\label{sec:3-D}
In the region (ii) of CP (i.e., $H_{fmax}<y<H_{Amax}$), we found the Spatial Velocity Correlation Function (SVCF) $C_{v_xv_x}(R)=\langle v_{x}(x+R)v_{x}(x)\rangle $ takes positive value for all relative displacements $R(0<R<L_x/2)$ (e.g., $\Lambda=175.4$ (CP) in Fig.~\ref{fig:10}).
Thus, the particles flow collectively in the same horizontal direction.

In Fig.~\ref{fig:10}, SVCFs at the height $H_{Amax}$ in LFP and GTP are also shown.
In LFP, the negative minimum of SVCF becomes larger when $\Lambda$ is increased, which means that the size of fluidized hole grows large.
In GTP (i.e., $\Lambda > 357.1$), there is no negative minimum in $C_{v_xv_x}(R)$ for all $R$ less than half of the system size($L_x/2$). 
This means the size of fluidized hole grows larger than the half of system size.

\subsection{Displacement of particles}
\label{sec:3-E}
In Fig.~\ref{fig:11}, the time-dependence of displacement vectors for each particle from the initial position are shown at $t/\tau = 100, 200, 300, 400$ when $\Lambda = 25.0$.
The displacement vectors from the dark region indicate that the direction of the displacement is right.
This is a clear evidence of the existence of the domain of collective particle motion.
The feedback of collective motion can clearly be seen during the time $t/\tau=200$ $\sim$ $t/\tau=400$.

In Fig.~\ref{fig:12}(a), we also show a snapshot of the displacement vectors for each particle at $t/\tau =150$ in $\Lambda = 175.4$.
The domain size of collective motion becomes smaller than that of Fig.~\ref{fig:11}.
Since the particle has a room to move freely compared with a weakly exited state (e.g., $\Lambda = 25.0$), the convective motion appears near the critical point $\Lambda_{CL}$.
This convectional mode is similar to the granular vibrated bed.

In LFP (Fig.~\ref{fig:12}(b)), particles at the region of broken layers behave such as vortex-like stable convection.
The most of excited particles drops on the condensed layer near the broken region.

In GTP (Fig.~\ref{fig:12}(c)), the particles at the broken layer are highly excited, which can reach an another side of the condensed layer through the periodic boundary condition.
This motion must influence the results on the velocity PDF and spatial velocity correlation function.

\subsection{Horizontal averaged particle current in CP}
\label{sec:3-F}
Since the excluded volume effect is dominant in CP, the collective motion of particle appears.
Figure~\ref{fig:11} shows that the displacement of particles in CP move in the same direction during the short time scale.
We can also confirm the above result quantitatively that the averaged particle current field summed up in the packed region (ii), 

\begin{equation}
{\bf j}({\bf x}_g, \Delta t) = \frac{1}{N_s} \sum_{\rm region (ii)} 
\sum_i {\bf v}_i (\Delta t) \delta ({\bf x}_i (\Delta t)-{\bf x}_g ),
\end{equation}

\noindent
where $N_s$ is the sample number of particles, ${\bf x}_g$ is grid with the width of the disk diameter $d$ and $\Delta t = N/(Collision Rate)$.
Figure~\ref{fig:13} shows that the time-evolution of the horizontal averaged particle current for one particle in $\Lambda = 25.0$ and $\Lambda = 175.4$.
In a relatively short time scale, the particle currents in the packed region take positive value. 
However, we found the currents fluctuate around zero in a long time scale.
The feedback time scale of the current becomes longer when $\Lambda$ is near the critical point.
Therefore, the collective motion appears strongly when the system is in the weakly excited state.

\subsection{Surface wave in CP}
\label{sec:3-G}
In CP, the dense layer packing is formed and it is stable in time because excitation is weak and the potential energy is dominant.
However, it is apparent that collective motion appears on the surface of the condensed layer.
What is the mechanism of the surface wave-like motion in CP?
There must be no surface tension in the macroscopic dissipative system, therefore the restoring force should originate from the balance between the gravity and the excitation, but it is not clear if it could be defined as an ordinary gravitational wave in fluid.

To study the dynamical behavior of surface wave in detail, simulations were done on the driving intensity $\Lambda$ varied from 25.0 to 188.6.
We performed simulations until the time $T=10240\tau$ for each $\Lambda$.
Figures~\ref{fig:14}(a) and (b) show the spatio-temporal pattern of surface height of the condensed layer in CP far from the CP-LFP critical point ((a) $\Lambda=25.0$ and (b) $100.0$).
And Fig.~\ref{fig:14}(c) shows the pattern near the CP-LFP critical point ((c) $\Lambda=188.6$).
The horizontal and vertical axes correspond to the space expansion ($x=0 \sim L_x$) and the time evolution ($T=0\sim 2000\tau$), respectively.
There are many domains in the spatio-temporal patterns, which represent the collective motion of surface appears.
The typical domain sizes, which are estimated by Fig.~\ref{fig:14}(c), are about $\Delta x \sim L_x/2$ and $\Delta t\sim 850\tau$.
These are correspond to $(k(=2\pi/(L_x/2)),\omega(=2\pi/850))=(0.0628,0.00739)$, respectively.
To study this collective motion quantitatively, we calculate the dynamical structure factor of the surface wave for each run.
The dynamical structure factor $S(k,\omega)$ is described as

\begin{eqnarray}
S(k,\omega) & = & \frac{1}{2\pi N_{gx}}\int_{-T}^T \sum_{i,j} 
\langle h_y(x_i,t)h_y(x_j,0) \rangle_T \nonumber \\
 & \times & \exp{(-ik(x_i-x_j))}\exp{(i\omega t)}dt,
\end{eqnarray}

\noindent
where $h_y(x,t)$ is the surface height of condensed layer; $N_{gx}$($=200$) is the total number of grids with the width of disk diameter $d$ in the horizontal direction and the discrete positions are taken by $x_i=(L_x/N_{gx})/2+(i-1)(L_x/N_{gx}), (i=1...N_{gx})$.
When $\Lambda = 188.6$, the strongest peak of dynamical structure factor in the $k-\omega$ space appeared at ($k,\omega) \sim (2\pi/(L_x/2), 0.0065$).
Thus, we confirmed these values were consistent with the estimation of the domain sizes in the spatio-temporal pattern.
When $\Lambda$ becomes low, the domain size in the vertical direction becomes small (Fig.~\ref{fig:14}).
On the other hand, the domain size in the horizontal direction is not changed very much.
This means that the periodical cycle of surface wave becomes shorter in low $\Lambda$.
These results are also confirmed by the computation of the dynamical structure factor for each $\Lambda$.
At a result, when the strongest peak ($k_{max},\omega_{max}$) was plotted in $k-\omega$ space for each $\Lambda$, $k_{max}$ stays around $k=2\pi/(L_x/2)$ and increases slightly upon decreasing $\Lambda$, but $\omega_{max}$ becomes large as $\Lambda$ go to 0.

We conclude the periodic behavior of the surface wave in CP shows the critical slowing down when $\Lambda \rightarrow \Lambda_{CL}$.
Note that the value of $\omega$ at the critical point takes a certain finite value, not $0$. 
It can be understood that the transition occurs before the break down of the condensed layers, which means the transition seems to be subcritical.
We also found that the amplitude of the surface fluctuation becomes large near the transition point $\Lambda_{CL}$. 
This is an interesting point that the system shows large fluctuation around the transition like the critical phenomena in equilibrium.
 
\subsection{Dynamics of fluidized holes}
\label{sec:3-H}
The localized excitation in LFP should resemble a circle in 3D system and reminds us of an oscillon\cite{umbanhowar_1996}, but they are different; the external vibration frequency is essential for the oscillon dynamics, but the localized excitation here does not have such a characteristic frequency.
In the large $N_w$ simulation, we can observe the merging process of two or more excitations, but their mode of interaction is not clear.
An interesting question here is if the transition from LFP to CP and/or the transition to GTP can be understood in terms of the local excitation; if LFP transform to CP when the distance between the excitations diverges?
Does LFP transform to GTP at the point where they condense?

To investigate the excitation dynamics, we calculate the density fluctuation for each position with the width of particle diameter $d$ in the system.
The driving intensity $\Lambda$ was varied from 200.0 to 476.2, in which the systems are in LFP and GTP.
We performed simulations until the time $T=10240 \tau$ for each $\Lambda$.
The particle number in each column, which is summed up from the bottom to the top, defines the density.
If the particle number is less than the averaged layer number $N_h$, we regard the column as the fluidized holes (black region in Fig.~\ref{fig:15}).
If the column has the particles more than the averaged layer number $N_h$, these columns are regarded as condensed layers.
This criterion seems not to be a rigorous definition of fluidized holes and condensed layers, however, we confirmed that the fluidized holes were distinguished very well.

In Fig.~\ref{fig:15}, the spatio-temporal patterns of the density fluctuation are shown in LFP ($N_w=200$; (a) $\Lambda=200.0$ and (b) $250.0$) and GTP ($N_w=200$; (c) $476.2$).
Near the critical point $\Lambda_{CL}$, the lifetime of a fluidized hole does not continue through the simulation (Fig.~\ref{fig:15}(a)).
When $\Lambda$ is increased, the region of a fluidized hole increases (Fig.~\ref{fig:15}(b)).
These results are consistent with the increase of the weight of the broader distribution $p$ for the fitting functions of velocity PDF (Sec.~\ref{sec:3-B-6}).
Above LFP-GTP transition point (Fig.~\ref{fig:15}(c)), the local excitation domain fluctuates largely and sometimes expands to the length of the system.
We found that the spatio-temporal pattern of a fluidized hole changes significantly through the transition from LFP to GTP.

\section{DISCUSSION}
\label{sec:4-0}

\subsection{Size dependence}
\label{sec:4-A}
In this subsection, we discuss the changes of the statistical properties in IHSHG model upon increasing the system size $N_w$.

We performed the almost same simulations described by Sec.~\ref{sec:3-0} for the systems with larger $N_w$, which means the aspect ratio in the system is changed.
To examine the size dependence, the numerical simulations for various $\Lambda$ with $N_w=400$ (i.e., $N=8,000$) and $N_w=1,000$ (i.e., $N=20,000$) are performed.
The simulation lengths for these data are 30,000 ($N_w=400$) and 10,000 ($N_w=1,000$) collisions per particle, respectively.
We found that the $N_w$ dependencies of the maximum packing fraction and the excitation ratio do not change when $N_w$ is increased.
However, the $N_w$ dependence of the flatness at the height $H_{Amax}$ clearly becomes larger in LFP and GTP.

In Fig.~\ref{fig:16}, the velocity PDFs $\phi(v)$  with fitting function in $N_w=200$ and $N_w=400$ are shown when $\Lambda = 250.0$ (LFP).
The difference of velocity PDF between $N_w=200$ and $N_w=400$ is very small.
However, the flatness becomes clearly larger when $N_w$ is increased (e.g., $N_w=200, f\sim 19.0$; $N_w=400, f\sim 23.0$).
We change $\beta$ for $N_w=400 (\beta=0.82)$ to fit slightly broader distribution than that of $N_w=200 (\beta=0.85)$.
The other fitting parameters for $N_w=400$ remain the same as in $N_w=200$, i.e. $\sigma_G=0.048, \tilde x_0=1.7$, and $p=0.25$ (Sec.~\ref{sec:3-B-6}).
The fitting function seems good agreement with the simulation data.
Using these parameters, we calculated the flatness value by using Eq.~(\ref{eqn:flatness}).
In the inset of Fig.~\ref{fig:16}, the flatness values both $\beta=0.85$ and $\beta=0.82$ are plotted as functions of the weight $p$.
We found that the flatness values with $\beta=0.82$ are larger than that with $\beta=0.85$.
This is consistent with the fact that the flatness in $N_w=400$ takes larger value than that in $N_w=200$.
 
Although there is one hole in the system when $N_w=200$ (Sec.~\ref{sec:3-0}), there should be more holes in the limit $N_w \rightarrow \infty $ to reach finite density of holes.
What is the average density of the holes?
It might have some relationship to the width of the surface wave in CP (Sec.~\ref{sec:3-G}).
In Fig.~\ref{fig:17}, the spatio-temporal pattern of the density fluctuations are shown in LFP ($N_w=400$; (a) $\Lambda=200.0$, (b) $250.0$, and (c) $333.3$) and GTP ((d) $\Lambda=454.5$, (e) $588.2$, and (f) $833.3$), respectively.
Near the critical point $\Lambda_{CL} (\Lambda=200.0)$, the lifetime of a fluidized hole is shorter than the simulation length (Fig.~\ref{fig:17}(a)), which is similar situation in the case of $N_w=200$. 
When $\Lambda$ is increased, the width of the holes becomes wider.
However, the system seems to contain two fluidized holes (Fig.~\ref{fig:17}(b),(c)), which is the different situation if it is compared with the case of $N_w=200$.
In $N_w=1000$, we confirmed the fact that there are multiple holes in the system. 
As mentioned above, since the maximum packing fraction for each $\Lambda$ is not dependent on the system size, the ratios of the total hole region to the system size for each $N_w$ are almost equal.
However, the average size of hole is not dependent on system size, since
it is determined by the driving intensity $\Lambda$.
At the result, the systems with large $N_w$ have several holes.
Above the LFP-GTP transition point ($\Lambda > 357.1$), the local excitation domain largely fluctuates (Fig.~\ref{fig:17}(d),(e)).
We can see the high-speed motion of the high density cluster when $\Lambda =833.3$ (Fig.~\ref{fig:17}(f)). 
When $N_w$ is increased, since there are many fluidized holes in the system, the fluctuation of hole width might become larger. 
In this situation, velocity PDFs for each hole have different variances for the width of holes, which means various broader distributions of horizontal velocity coexist with in the system. 
In general, the flatness takes large value when the spatial heterogeneity become large.
This spatial heterogeneity of the velocity PDFs at fluidized holes might be the reason why the flatness takes larger value when $N_w$ is increased.  

\subsection{CP-LFP transition}
\label{sec:4-B}
In this subsection, we summarize and discuss the results of numerical simulation about the transition from CP to LFP.

We found that the value of the maximum flatness at $y=H_{fmax}$ shows a sharp rise around the transition point between CP and LFP ($\Lambda_{CL}\sim 188.7$). 
The flatness $f$ becomes larger than $20$, and we demonstrated that this unexpected large value of $f$ comes from the fact that the velocity PDF in LFP consists of two components; the narrow Gaussian distribution that comes from the dense layer and the broad stretched exponential distribution in the fluidized holes (Sec.~\ref{sec:3-B-6}).
Actually, velocity PDF $\phi(v)$ in LFP can be expressed very well by the sum of the narrow Gaussian $\phi_G(v)$ and the broader stretched exponential $\phi_S(v)$ as $\phi(v) = p \phi_{S}(v) + (1-p) \phi_{G}(v)$, with the weight $p\quad (0\le p\le 1)$.
Therefore, CP and LFP are distinguished by the existence of the fluidized holes in the latter.

Near the critical point $\Lambda_{CL}$ in LFP ($188.7 < \Lambda < 204.1$), we found that the lifetime of a fluidized hole is shorter than the simulation length (Fig.~\ref{fig:15}(a) and Fig.~\ref{fig:17}(a)).
On the other hand, for $204.1 < \Lambda < 357.1$, fluidized holes persist in time. 
In LFP, the majority of the particles are in the condensed layer, and it is punctured by the fluidized holes, from which the high-speed particles are blown.
We observed the stable flow of particles between the fluidized holes and the edges of the condensed layer (Fig.~\ref{fig:12}(b)), and this stable flow stabilizes the isolated fluidized hole in the condensed layer.
The horizontal velocity correlation between the opposite edges of the holes becomes negative, and this should be the origin of the negative minimum in the spatial velocity correlation function (Fig.~\ref{fig:10}).
We observed the size of the holes gets larger when $\Lambda$ is increased.
This result can be understood by the fact that the size is determined by the distance that the blown up particles reach, which is, in turn, determined by the driving intensity $\Lambda$.

The time scale of the surface fluctuation in CP becomes larger as $\Lambda$ approaches the critical point $\Lambda_{CL}$ from below.
On the other hand, the length scale seems to stays fairly constant.
From the detailed analysis of the dynamical structure factor of surface wave, the typical frequency $\omega$ does not vanish at the critical point $\Lambda_{CL}$, which suggests the transition between CP and LFP is subcritical.

\subsection{LFP-GTP transition}
\label{sec:4-C}
As the driving intensity is increased, the density and the size of the fluidized hole increase, and at $\Lambda_{LG} \sim 357.1$ neighboring holes start interfering with each other.
At the transition point $\Lambda_{LG}$, the blown up particles at one hole barely reach the neighboring holes.
As we have discussed, the distance that the flight of blown up particles also determines the size of the holes, therefore, at $\Lambda=\Lambda_{LG}$, we expect the half of the condensed layer structure should be destroyed by the fluidized holes.
In Fig.~\ref{fig:1}, we found that the averaged packing fraction at $\Lambda=357.1$ is $A \sim 0.48$, which is a little larger than the half of that of the value of the closed packing fraction ($A_{CP} \sim 0.901$) in 2D.

In the system with $N_w=200$, for which most of the simulations were done, there appears only one fluidized hole.
In this situation, the interference between holes takes place only through the periodic boundary.
Therefore, the LFP-GTP transition occurs when both the distance and the size of holes become a half of the system size $L_x/2$.
And the spatial velocity correlation function $C_{v_xv_x}(R)$ between the opposite edges of the hole becomes negative.
In Fig~\ref{fig:10}, we actually confirm the results on the negative value of SVCF $C_{v_xv_x}(R=L_x/2)$ when $\Lambda > \Lambda_{LG}$.

Another peculiar aspect of the system with $N_w=200$ is that the weight of the stretched exponential distribution in velocity PDF (i.e., $p$) stays $\sim 0.5$ for a finite range of $\Lambda$ in GTP (Fig.~\ref{fig:7}).
This plateau behavior may be also interpreted as the feedback effect of the periodic boundary condition which enforces the number of holes to be one for a certain parameter region of $\Lambda$.
In the larger systems with $N_w=400$ and $1,000$, we observed more than one holes in the system, and found two different holes interfere with each other in GTP.

The interference between holes destroys the stable particle flow within an isolated fluidized hole (Fig.~\ref{fig:12}(c)), and the holes are destabilized dynamically (Fig.~\ref{fig:15} and Fig.~\ref{fig:17}).
In Fig.~\ref{fig:15}(c) and Fig.~\ref{fig:17}(d),(e),(f), which are the spatio-temporal patterns in GTP, we found many high-speed moving small clusters.
These clusters divide the region of the fluidized holes (i.e., black region) into several parts.
In the patterns of LFP (Fig.~\ref{fig:15}(b) and Fig.~\ref{fig:17}(b),(c)), there are no such moving clusters.
Therefore, these high-speed moving small clusters are thought as the peculiar phenomena in GTP.
When these clusters move across the fluidized holes in the spatio-temporal patterns, the lifetime of the fluidized holes becomes shorter.
Although there is still room for the definition of quantitative values to distinguish the phases between LFP and GTP, we found that the maximum lifetime of fluidized holes in the spatio-temporal patterns as a function of $\Lambda$ changes drastically between LFP and GTP.
Thus, we conclude that the LFP-GTP transition is interpreted as the dynamical destabilization transition of the fluidized holes.

The particle dynamics in GTP are very different from that of LFP.
The average density is quite low, but it is very different from the ordinary molecular gas phase.
The density fluctuation is very large and this fluctuation causes turbulent motion due to the gravity.
Even for very large $\Lambda$, the high-density clusters which look like the condensed layer are formed temporally.
It might be also different from ordinary fluid turbulence.
In the finite system with finite height, turbulence may disappear when the driving $\Lambda$ is large enough, but in the system with infinite height, it appears that turbulent motion persists however large the driving $\Lambda$ is.

\subsection{Various limits of IHSHG model}
\label{sec:4-D}
The dynamics of IHSHG model is very different from those of not only an ordinary molecular system but also of a conventional granular vibrated bed.
Let us discuss the relationship between the present model and the vibrated bed, where the system is driven by a vibrating plate at a given frequency.
In such a case, the amplitude $a$ and the angular frequency $\omega$ of the vibration are two independent parameters, but the control parameter is usually taken as the ratio of the accelerations $\Gamma\equiv a\omega^2/g$.
In most experimental situations, the external frequency determine the time scale of the system and the collision interval is directly related to $1/\omega$.
In the present case, the ratio of the energies $\Lambda =k_BT_w/mgd$ is the only control parameter and no external time scale is imposed.
This situation may correspond to that in the vibrated bed when $1/\omega$ is much shorter than any other of the relevant time scales in the system.
                            
Since we fixed the parameters of IHSHG model at $N_h=20$ and $r=0.9$, the dependence of the behavior by changing $N_h$ or $r$ is not obvious.
When the restitution coefficient is fixed at unity, we can obtain the density profile of point particles (i.e., the diameter $d \rightarrow 0$) within the framework of the statistical mechanics.
Under the situation where the temperature and the gravity are uniform in the vertical direction, the density profile as a function of the height $y$ is exponentially decreasing function $\sim \exp{(-mgy/k_BT)}$ and the excitation ratio $\mu$ is always unity.
Since the point particles can be compressed infinitely, the profile simply shifts.
Therefore, no phase changes occur in the system when the driving intensity is changed.
However, in the case of finite diameter of particle $d$, which cannot be compressed infinitely, the density profile becomes different from that of point particles when the driving intensity becomes relatively low.
Under a certain intensity of driving force, a fraction of hard particles condenses from the bottom toward the surface.
Within the framework of Enskog theory\cite{hong_1999}, the density profile for hard spheres remains finite at the close packed density near the bottom and the exciation ratio $\mu$ becomes less than unity.
It is difficult to predict the system behavior when $N_h$ is changed, because $N_h$ dependence of the system might be significantly changed for various restitution coefficient $r$.
However, in the case of $N_h=1$, the system is almost the same situation as 1D model heated from the bulk (Sec.~\ref{sec:2-A-2}).
Systematic studies of the system behaviors by changing these parameters will be the subject of future work.

\section{CONCLUDING REMARKS}
\label{sec:5-0}
In this paper, we performed numerical simulations for a two-dimensional inelastic hard disk system heated from the boundaries under gravity (IHSHG). 
Upon increasing the heat bath temperature, the system exhibits three distinct phases, namely, the condensed phase, the locally fluidized phase, and the granular turbulent phase.
In the condensed phase, most of the particles are aggregated around the bottom of the system with an almost close packed density and the state is fairly stable; this is because the excitation of the heat bath is weak and the potential energy (gravity) is dominant.
The collective motion appears on the surface and in the bulk of the condensed layer.
In the locally fluidized phase, the dense layer packing is locally broken by the excitation of the heat bath.
The high-speed particles are blown upward from the holes, which are the broken layer.
However, the location of holes is fairly stable.
In the granular turbulent phase, the motion of particles becomes very active and the kinetic energy is dominant, because the most of particles are highly excited by the heat bath.
The location of holes becomes unstable in time, and the condensed layer appears only temporally.
Over the whole region, the flatness $f(y=H_{fmax})$ is different from 3, which is the value for the Gaussian distribution, but it is remarkable that $f(y=H_{fmax})$ becomes very large in the locally fluidized phase.
The transition from the condensed phase to the locally fluidized phase is distinguished by the existence of fluidized holes.
On the other hand the transition from the locally fluidized phase to the granular turbulent phase is interpreted as the destabilization transition of the fluidized hole.

In the study of the dissipative structure system, such as Rayleigh-B\'enard convection, it is well known that the several phase changes (bifurcations) occur when the external driving is increased.
However, in dissipative discrete element (granular) systems, there are very few studies so far.
We believe our study will offer a step for understanding the macroscopic universal characters in the studies of fluidization of dissipative (inelastic) discrete element system with a heat bath under gravity and constructing the non-equilibrium statistical mechanics.

\acknowledgments

I would like to thank to Professor H.~Nakanishi for making helpful discussions.
I also wish to thank Professor T.~Odagaki and Professor H.~Kawai for valuable suggestions.
And I acknowledge helpful discussion with Professor H.~Hayakawa on granular material.
This work is published as the author's Ph.D Thesis\cite{isobe_thesis} in Kyushu University.
A part of the computation in this work was done by the facilities of the Supercomputer Center, Institute for Solid State Physics, University of Tokyo.

%
%
%
%

\begin{figure}
\caption{
The maximum packing fraction $A_{max}$ and the excitation ratio $\mu$ vs. the driving intensity $\Lambda$.
The other parameters are fixed at $(r, N_w, N_h)=(0.9,200,20)$.
}
\label{fig:1}
\end{figure}

\begin{figure}
\caption{
A series of snapshots (interval time is $100\tau$) for the three typical values of $\Lambda$, a1---a3 ($\Lambda=181.8$), b1---b3 ($\Lambda=250.0$), c1---c3 ($\Lambda=666.7$).
The other parameters are fixed at $(r,N_w,N_h)=(0.9,200,20)$.
}
\label{fig:2}
\end{figure}

\begin{figure}
\caption{
The collision rate $C_{total}/T_{total}$ vs. the driving intensity $\Lambda$.
Exponential fitting curves for both LFP and GTP regime are also shown. Each regime has a different exponent.
}
\label{fig:3}
\end{figure}

\begin{figure}
\caption{
The horizontal velocity PDFs for the three typical $\Lambda$s (CP:$\Lambda=181.8$, LFP:$\Lambda=250.0$, GTP:$\Lambda=666.7$).
The other parameters are fixed at $(r,N_w,N_h)=(0.9,200,20)$.
An inset in the upper left-hand corner shows the Gaussian nature of the central regions for each distribution function.
}        
\label{fig:4}
\end{figure}

\begin{figure}
\caption{
The flatness $f(H_{fmax})$ vs. the driving intensity $\Lambda$.
The other parameters are fixed at $(r,N_w,N_h)=(0.9,200,20)$.
The values for Gaussian distribution $(f=3)$ and the exponential distribution $(f=6)$ are indicated in the figure.
}        
\label{fig:5}
\end{figure}

\begin{figure}
\caption{
The velocity PDFs and their fitting functions in LFP and GTP are shown.
The data are scaled by the standard deviation $\sigma$.
Open circles and filled squares denote that PDFs in $\Lambda=250.0$ with the parameters, $x_0/\sigma_G=1.7$, $\beta=0.85$, and $p=0.25$,
and $\Lambda=476.2$ with the parameters, $x_0/\sigma_G=1.7$, $\beta=0.85$, and $p=0.49$, respectively.
}
\label{fig:6}
\end{figure}

\begin{figure}
\caption{
The values of weight $p$ (the ratio of Gaussian to exponential behavior) for each $\Lambda$ are shown.
}
\label{fig:7}
\end{figure}

\begin{figure}
\caption{
The height of maximum values of both packing fraction and flatness for each $\Lambda$ are shown.
}
\label{fig:8}
\end{figure}

\begin{figure}
\caption{
Both maximum packing fractions and packing fractions at the height of $H_{fmax}$ for each $\Lambda$ in CP are shown.
}
\label{fig:9}
\end{figure}

\begin{figure}
\caption{
Spatial velocity correlation functions at the height of $H_{Amax}$ in the horizontal direction for each $\Lambda$ vs. the relative displacement $R$.
}
\label{fig:10}
\end{figure}

\begin{figure}
\caption{
The time-dependence of displacement vectors for each particle from the initial position are shown at $t/\tau = 100, 200, 300, 400$ in $\Lambda = 25.0$. 
The displacement vectors from the dark regions indicate that the particles move toward right-hand side from the initial position.
}
\label{fig:11}
\end{figure}

\begin{figure}
\caption{
(a) The displacement vectors for each particle from the initial position are shown at $t/\tau = 150$ in $\Lambda = 175.4$, (b) at $t/\tau = 50$ in $\Lambda = 250.0$, and (c) at $t/\tau = 50$ in $\Lambda = 400.0$. 
The displacement vectors from the dark regions indicate that the particles move toward right-hand side from the initial position.
}
\label{fig:12}
\end{figure}

\begin{figure}
\caption{
The time-dependence of the horizontal averaged particle current for one particle in CP is shown.
}
\label{fig:13}
\end{figure}

\begin{figure}
\caption{
The spatio-temporal patterns of a surface height of the condensed layer far from the critical point ((a) $\Lambda=25.0$ and (b) $100.0$) and near the critical point ((c) $\Lambda=188.6$) are shown.
The horizontal and vertical axes correspond to the space expansion ($x=0 \sim L_x$) and the time evolution ($T=0\sim 2000\tau$), respectively.
}
\label{fig:14}
\end{figure}

\begin{figure}
\caption{
The spatio-temporal patterns of the density fluctuations are shown in LFP and GTP.
($N_w=200$; (a) $\Lambda=200.0$, (b) $250.0$, and (c) $454.5$).
The horizontal and vertical axes correspond to the space expansion ($x=0 \sim L_x$) and the time evolution ($T=0\sim 10240\tau$), respectively.
The black region corresponds to the fluidized holes.
}
\label{fig:15}
\end{figure}

\begin{figure}
\caption{
The velocity PDFs with fitting function in both $N_w=200$ and $N_w=400$ 
are shown ($\Lambda=250.0$).
The data are scaled by the standard deviation $\sigma$.
An inset in the upper right-hand corner shows the calculated flatness values as a functions of the weight $p$ for both fitting parameters (i.e., $\beta=0.85$ and $\beta=0.80$) by using Eq.~\ref{eqn:flatness}.
}
\label{fig:16}
\end{figure}

\begin{figure}
\caption{
The spatio-temporal patterns of the density fluctuations are shown in LFP
($N_w=400$; (a) $\Lambda=200.0$, (b) $250.0$, and (c) $333.3$ )
 and GTP ((d) $\Lambda=454.5$, (e) $588.2$, and (f) $833.3$ ).
The horizontal and vertical axes correspond to the space expansion 
($x=0 \sim L_x$) and the time evolution ($T=0\sim 10240\tau$), respectively.
The black region corresponds to the fluidized holes.
}
\label{fig:17}
\end{figure}

\end{document}